\documentclass[conference]{IEEEtran}
\pdfoutput=1
\usepackage[utf8]{inputenc}
\usepackage{graphicx}
\usepackage{color}
\usepackage{algorithm}
\usepackage{algpseudocode}
\usepackage{amsmath}
\usepackage{graphicx}
\usepackage{caption}
\usepackage{float}
\usepackage{graphics}
\usepackage{cite}
\usepackage{multirow}
\usepackage{xcolor,colortbl}
\usepackage{varwidth}
\usepackage{hhline}
\usepackage{subfigure}
\usepackage{adjustbox}
\usepackage{tabularx}

\newcolumntype{Y}{>{\centering\arraybackslash}X}

\newcommand{\vars}{\texttt}
\newcommand{\func}{\textrm}
\let\oldReturn\Return
\renewcommand{\Return}{\State\oldReturn}

\title{Fast End-to-End Integrity Verification for High-Speed File Transfers}
\date{April 2018}

\begin{document}

\author{\IEEEauthorblockN{Engin Arslan and Ahmed Alhussen}
\IEEEauthorblockA{Computer Science and Engineering, University of Nevada, Reno\\
%Reno, NV\\
Email: earslan@unr.edu, aalhussen@nevada.unr.edu}
}

\newcommand{\algoName}{FIVER}

\maketitle

\begin{abstract}
The amount of data generated by scientific and commercial applications is growing at an ever-increasing pace. This data is often moved between geographically distributed sites for various purposes such as collaboration and backup which has led to significant increase in data transfer rates. Surge in data transfer rates when combined with proliferation of scientific applications that cannot tolerate data corruption triggered enhanced integrity verification techniques to be developed. End-to-end integrity verification minimizes the likelihood of silent data corruption by comparing checksum of files at source and destination servers using secure hash algorithms such as MD5 and SHA1. However, it imposes significant performance penalty due to overhead of checksum computation. In this paper, we propose Fast Integrity VERification (\algoName) algorithm which overlaps checksum computation and data transfer operations of files to minimize the cost of integrity verification. Extensive experiments show that \algoName~is able to bring down the cost from 60\% by the state-of-the-art solutions to below 10\% by concurrently executing transfer and checksum operations and enabling file I/O share between them. We also implemented \algoName-Hybrid to mimic disk access patterns of sequential integrity verification approach to capture possible data corruption that may occur during file write operations which \algoName~may miss. Results show that \algoName-Hybrid is able to reduce execution time by 20\% compared to sequential approach without compromising the reliability of integrity verification.

\end{abstract}

\section{Introduction}
According to a study by Forrester Research~\cite{forrester10}, 77\% of the 106 large IT organizations operate three or more datacenters and run regular backup and replication services among these sites. More than half of these organizations have over a petabyte of data in their primary datacenter and expect their inter-datacenter throughput requirements to double or triple over the next couple of years. There has been a very similar trend in scientific applications for the last decade as large scientific experiments such as environmental and coastal hazard prediction~\cite{Klein200335}, climate modeling~\cite{Climate}, genome mapping~\cite{BLAST}, and high-energy physics simulations~\cite{CMS, ATLAS} generate data volumes reaching petabytes per year. This massive amount of data often needs to be moved for various purposes such as processing, collaboration, and archival. While most of earlier works focused on optimization of data transfers~\cite{europar13,sc16,TCC_2016,raj-ccgrid14}, the integrity of transfers are also critical for many applications such as Dark Energy Survey~\cite{des} and Sky Survey Simulation~\cite{habib2016hacc} as they rely on correctness of data to operate. 

On the other hand, as data transfer rates are rapidly increasing, traditional integrity verification mechanisms fall short to detect corruption. For example, some of the components of data transfers have built-in integrity verification mechanisms, however they are either weak or applicable to subset of available systems. For example, TCP checksum fails to detect errors once in 16 million to 10 billion packets~\cite{stone2000crc}. As a result, researchers observed up to 5\% data corruption for file transfers in 100 Gbps network~\cite{kettimuthu2018transferring}.

End-to-end integrity verification is introduced to reduce the likelihood of accepting corrupted data. The basic implementation of end-to-end integrity verification for data transfers works as follows: Sender first reads the file from storage and sends it to destination. Once data transfer is completed, the sender reads the file again to compute checksum using a hash algorithm such as MD5 and SHA1. Receiver does the same and computes the checksum after file is fully received and written to the storage. Finally, the sender and receiver exchange the computed checksum values and compare. If checksum values of source and destination servers are the same, then the transfer is marked as successful, otherwise the file at destination server is assumed to be corrupted and transfer is restarted. If the dataset consists of multiple files, then the transfer of next file will begin only after previous file's integrity verification is completed successfully.

While end-to-end integrity verification is crucial for many applications, it imposes significant overhead since checksum computation would take long time, especially for large files or busy servers. As opposed to basic approach which runs file transfers and checksum computation sequentially for every file, Globus~\cite{globus} supports file-level pipelining in which checksum of a file can be calculated while another file is being transferred. However, it fails to overcome performance issues due to two main reasons. First, it is hard to achieve perfect pipelining of data transfers and checksum computation if the file sizes in dataset are different or the speed of transfer and checksum computation operations is different. Secondly, running checksum computation and transfer operations of two different files simultaneously may cause I/O contention due to which both processes may experience slow down. Liu et al.~\cite{liu2016towards} proposed block-level pipelining to address the limitations of the file-level pipelining by dividing large files into smaller blocks to better overlap transfers and checksum operations for datasets with mixed file sizes. However, it requires an upfront work to determine the block size that achieves optimal pipelining for different network and host configurations. Moreover, dividing large files into smaller blocks could deteriorate transfer throughput when network transfer runs faster than checksum computation and has to stay idle while checksum calculation is going on which may trigger TCP window size reset for every block transfer~\cite{RFC2581}.

In this paper, we propose Fast Integrity VERification (\algoName) that runs transfer and checksum operations of a file simultaneously to reduce I/O overhead and achieve better pipelining. While previous works focus on overlapping checksum computation and transfer operation of different files or blocks, \algoName~overlaps those operations for the same file. For example, if checksum computation of a file takes 30 seconds and transfer time takes 10 seconds, then \algoName~finishes transfer and integrity verification in around 30 seconds with the help of simultaneous execution of both processes. Secondly, as opposed to reading each file from storage twice (one for file transfer and one for checksum computation), \algoName~reads the files once and share I/O between checksum and transfer operations since the second read would end up being served from cache memory anyway for files that are smaller than free memory space. For example, data transfer process reads $n$ bytes of the file and share it with checksum process before transferring it such that checksum process does not have to run system calls to read same data from page cache. Thus, \algoName~provides similar integrity guarantees as other existing approaches but provides significant performance benefits.

Contributions of this paper are as follows:
\begin{itemize}
    \item We introduce \algoName~that runs network transfer and checksum computation of a file concurrently to minimize the overhead of integrity verification.
    \item We enable file I/O sharing between transfer and checksum operations to obviate system calls for checksum file I/O.
    \item We propose \algoName-Hybrid that uses \algoName~for small files and sequential approach for large files to reduce the cost of running integrity verification without compromising the reliability of integrity verification.
    \item We run extensive analysis to evaluate the performance of \algoName~and \algoName-Hybrid in various network settings using various datasets.
\end{itemize}

The rest of paper is organized as follows: Section~\ref{sec:related} briefly describes related work and Section~\ref{sec:system_desing} presents the motivation and design principles of proposed algorithms. We give experimental results in Section~\ref{sec:results} and conclude the paper with summary and future directions in Section~\ref{sec:conclusion}.

\section{Related Work}\label{sec:related}
 As increasing number of applications rely on the accuracy of data to perform properly, integrity verification has been widely studied in many areas including storage outsourcing~\cite{ateniese2007provable,zhu2012cooperative, liu2015external}, long term achieves~\cite{maniatis2005lockss,vigil2015integrity}, file systems~\cite{ma2014ffsck, zhang2013zettabyte,zhang2010end}, databases~\cite{arshad2018efficient}, provenance~\cite{hasan2009case} and data transfer~\cite{liu2016towards}. 

Zhang et al.~\cite{zhang2010end} evaluated Zetabyte Files System (ZFS) in terms of robustness to disk and memory fault injections. It has been found that while ZFS is able to detect and mostly recover from disk corruptions, it is susceptible to memory corruptions since it does not check the integrity of data blocks when they reside in the memory. \algoName~also makes similar assumption and rely on existing control mechanisms to recover from possible data corruption in memory. On the other hand, Error Correcting Codes (ECC)~\cite{chen1984error} has been proposed to detect and recover from the most single and multi-bit failures as well as memory leaks~\cite{qin2005safemem}. Meza et al.~\cite{meza2015revisiting} has showed that very large portion of memory errors (over 98\%) are correctable by the common ECC implementation. Yet,
more advanced error correction algorithms such as Chip-kill~\cite{dell1997white} are proposed to recover from more sophisticated (up to 4-adjacent bit corruption) errors.

Xiong et al.~\cite{xiong2016bloom} proposed \emph{fsum} that uses bloom filter compute the checksum of very large datasets stored in long term archival storage. Instead of calculating checksum for each file, they partition files into blocks such that multiple threads can process different portions of the file simultaneously. To achieve ordering among threads, bloom filter is used to combine the results from threads. Once all the blocks are processed and corresponding hash values are inserted into the bloom filter, final checksum is calculated by computing the hash of the bloom filter. \emph{fsum} runs up to 4x faster than traditional file-level checksum computation approach but comes with the cost of probability of false positive identification due to relying on bloom filter.

Globus~\cite{globus} supports end-to-end integrity verification for data transfers. It can pipeline data transfers and checksum computation to minimize the overhead of integrity verification. However, it calculates the checksum of files after their transfer completes. That is, files are read twice in the source server, one to send to destination and one to compute checksum. Yet, its pipelining approach fails to work well when dataset consist of mixed file sizes. Liu et al.~\cite{liu2016towards} proposed block-level pipelining to achieve better overlapping of checksum computation and file transfer operations. It reduces execution time considerably especially when dataset is composed of files with mixed sizes. Similar to Globus, block-level pipelining also reads files twice at source server. However, as opposed to Globus, its second read is faster for all file sizes since blocks are small enough to be kept in memory for some time, so once a block is read for transfer operation, it will be cached. When checksum computation process attempts to read the file block, it will find it in the memory. On the contrary, \algoName~reads files once and run the transfer and checksum computation processes simultaneously, reducing I/O overhead and time required to compute checksum.

On the other hand, previous work in high speed networks mostly focus on transfer scheduling~\cite{globus,ScienceCloud_2013,kosar04}, throughput optimization~\cite{arslan2018high, arslan2018big, raj-ccgrid14, TCC_2016}, and power consumption optimization~\cite{ismail-sc15}. Globus~\cite{globus} offers data transfer and sharing services and it is well adopted by research community. Similarly, Stork~\cite{kosar04} is proposed to handle data transfer tasks by integrating them into job schedulers such as Condor. HARP~\cite{sc16} models data transfers using historical data and real time sampling and uses this model to estimate the set of values for application layer transfer parameters that would maximize the throughput of given transfer tasks. PCP~\cite{TCC_2016} finds the optimal values for transfer parameters by running a series of sample transfers in the runtime. Alan et al.~\cite{ismail-sc15} proposed scheduling algorithms that can tune application layer transfer parameters to find a balance between transfer throughout and energy consumption at the end hosts. The algorithms monitor CPU usage of end hosts and estimates energy consumption with the help of models that relate CPU usage to energy consumption. Then, a cost function is used to determine the energy efficiency of each configuration based on transfer throughput and energy consumption values. Finally, a configuration with minimum cost function is identified and used in the rest of transfer.

%Results show that executing data transfer and checksum computation at the same time does not affect the performance of data transfer. Moreover it does compromise the validity of integrity verification in the source server since sharing I/O for checksum computation and transfer operations does not affect the fact that files are still read from disk. At the receiving end, \algoName~again overlaps transfer operations and checksum computation but does not share I/O accesses this time. Instead, checksum computation tries to read the file and it's being written to the disk. Although checksum computation 
%checksum computation attempts to it first receives data clocks from reads the data after they are received from socket and written to disk. While block level pipelining reads the data after certain size of it is received (e.g., after receiving 500MB if block size id 500 MB) the read operation is done is smaller sizes (typically 4KB).
%it reads files twice in the source host, one to data and one to compute checksum. 

\section{System Design of \algoName} \label{sec:system_desing}
The simple approach (sequential) implements integrity verification in three steps. In the first step, a file is transferred from source to destination using preferred transfer application. Once the transfer of the file is completed and it is written to the storage at destination, the second step starts during which checksum of original file at source and transferred copy at destination are computed using desired hash function such as MD5 or SHA1. In the third and final step, checksum values of original file and transferred copy are exchanged between source and destination servers to compare. If the checksum values are the same, then file transfer is marked as finished. Otherwise, transferred copy of the file will be assumed corrupted and all three steps are executed again.

\begin{table}
\begin{centering}
\resizebox{0.48\textwidth}{!}{
\begin{tabular}{ |c| c| c|}
\hline
{\bf Specs} & {\bf Source} &  {\bf Destination}\\
\hline
{\bf File System} & 24 HDDs RAID 0 & 12 SSDs RAID 0\\
\hline
{\bf CPU} &  \multicolumn{2}  {c|} {12 x Intel Xeon E5-2643 3.40GHz}\\
\hline
{\bf Memory (GB)} & \multicolumn{2} {c|} {16} \\
\hline
{\bf Bandwidth (Gbps)} &  \multicolumn{2} {c|} {100} \\
\hline
{\bf RTT (ms)} &  \multicolumn{2} {c|} {0.2 (LAN) \& 89 (WAN)}\\
\hline
\end{tabular}}
\caption{System specification of ESNet network.}\label{tab:esnet-lan}
\vspace{-3mm}
\end{centering}
\end{table}

\iffalse
\begin{figure*}[!ht]
\begin{center}
\subfigure[Source Server]{
\includegraphics[keepaspectratio=true,angle=0,width=88mm] {figures/motivating_sender}
\label{fig_ot_small}}
\hspace{-4mm}
\subfigure[Destination Server]{
\includegraphics[keepaspectratio=true,angle=0,width=88mm] {figures/motivating_receiver}
\label{fig_ot_large}}
\caption{Checksum computation process reads files from memory if file size is smaller than available memory space.}
\label{fig:motivating_example}
\end{center}
\end{figure*}
\fi

\subsection*{Motivating Example}

The main objective of running end-to-end integrity check is to detect possible data corruption by comparing the checksum of the file at the source and destination servers. On the other hand, operating systems are designed to minimize cache misses, so if a file is recently read or written, it will be kept in the memory to optimize successive accesses to the file. This causes checksum processes to be restricted to cached copy for small files even though checksum calculation is run after file transfers. That is, operating system of destination server will cache the whole file in memory as it is being streamed from network and written to disk if the file size is smaller than free memory space. So, when checksum process starts reading the file after its transfer is completed, the process will be served from memory because of which running checksum computation after file transfer does not ensure that the file will be read from disk. Similarly, files will be cached after they are read for transfer at source server. Thus, file I/O of checksum computation will be served from cache memory.

%If the checksum at destination server is computed while it is being written to the disk, operating systems will serve the file from memory since it will cache the file as it's being written. As a result, data corruptions that happen when pages of the memory are written to disk will not be covered by integrity verification which is not desirable. On the other hand, it is hard to guarantee that files will be read from disk when checksum computation is executed. 
\begin{figure}
\begin{center}
\includegraphics[keepaspectratio=true,angle=0,width=0.48\textwidth] {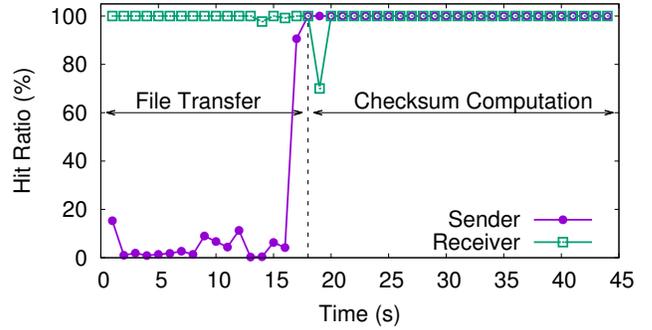}
\caption{Checksum computation processes at sender and receiver servers read files from memory when file sizes are smaller than free memory space.}
\label{fig:motivating_example}
\end{center}
\end{figure}

Figure~\ref{fig:motivating_example} illustrates the cache statistics of source and destination servers when sequential approach is employed for a transfer of a file with 8GB size. The system specification of test environment is given in Table~\ref{tab:esnet-lan}. The transfer of a file takes around 18 seconds and remaining 27 seconds are spent to calculate checksum. As shown in the figure, both sender and receiver servers have negligible cache misses and performs almost consistent 100\% cache hit ratios for checksum computation. Sender-side hit ratio is small during the file transfer since this is the first time the file is being read. Once file transfer is complete, it starts to read the file again to compute the checksum to exchange and verify the integrity of the file at destination. All file read operations for checksum computation are being served from memory (page cache), so hit ratio is 100\% throughout the process. On the other side, file transfer operation does not involve any file read I/O at the receiver server, as a result no cache misses are reported. Upon the completion of file transfer and disk write, the file is opened again for read to compute checksum which ends up being served from memory. 

Hence, cache hit ratio stays at 100\% since the pages of the file are kept in the memory after it is received and written. \emph{Thus, if free memory space is bigger than file size, both sender and receiver servers will end up operating on cached version of the file during checksum computation.} We observed that most production systems use Data Transfer Nodes with 64 GB or larger memory size which would cause file reads of most checksum calculations end up at the cache memory as average file size of scientific data transfers are in the order of megabytes~\cite{liu2018cross}. Although, one can clear the cache after file transfer, it requires root access and could have detrimental impact on other processes running on the servers, so it is neither possible nor desirable in most cases. Hence, we implemented \algoName~that runs transfer and computation operations simultaneously and shares I/O accesses to better utilize network and end system resources.

\begin{algorithm}
\scriptsize
\centering
\caption{--\algoName~Sender}
\begin{algorithmic}[1]
%\vspace{1mm}
\Statex
\State \textbf{global} \vars{queue, socket} \Comment Fixed size, synchronized queue
\Function{send}{\vars{fileName}, \vars{fileSize}}
	\State \vars{file} = \func{openForRead}(\vars{fileName})
	\State \Call{computeChecksum}{\vars{fileName}, \vars{fileSize}} \Comment Runs on a separate thread
	\While {\vars{file}.\func{read}(\vars{buffer}) $>$ 0}
	    \State \vars{socket}.\func{write}(\vars{buffer})
	    \State \vars{queue}.\func{add}(\vars{buffer}) \label{line:add}  \Comment Synchronized operation
	 \EndWhile
\EndFunction
\item[]
\Function{computeChecksum}{\vars{fileName}, \vars{fileSize}}
    \State MessageDigest \vars{md}
	\State \vars{totalBytes} $=$ 0
	\While {\vars{totalBytes} $<$ \vars{fileSize}}
	    \State \vars{buffer} = \vars{queue}.remove()    \Comment Synchronized operation
	    \State \vars{md}.update(\vars{buffer})
	    \State \vars{totalBytes} += \vars{buffer.length}
	 \EndWhile
	 \State \vars{localChecksum} $=$ \vars{md}.\func{digest()}
	 \State \vars{remoteChecksum} = \vars{socket}.\func{read}()
	 \If {\vars{localChecksum} $\neq$ \vars{remoteChecksum}}
	    \State \Call{Send}{\vars{fileName}, \vars{fileSize}}\Comment Transfer again if verification fails
	 \EndIf
\EndFunction
\end{algorithmic}
\label{alg:sender}
\end{algorithm}

\begin{algorithm}
\scriptsize
\centering
\caption{--\algoName~Receiver}
\begin{algorithmic}[1]
%\vspace{1mm}
\Statex
\State \textbf{global} \vars{queue, socket}\Comment Fixed size, synchronized queue
\Function{receive}{\vars{fileName}, \vars{fileSize}}
	\State \vars{file} = \func{openForWrite}(\vars{fileName})
	\State \Call{computeChecksum}{\vars{fileSize}} \Comment Runs on a separate thread
	\While {\vars{socket}.\func{read}(\vars{buffer}) $>$ 0}
	    \State \vars{file}.\func{write}(\vars{buffer})
	    \State \vars{queue}.\func{add}(\vars{buffer}) \label{line:add2}  \Comment Synchronized operation
	 \EndWhile
	
\EndFunction
\item[]
\Function{computeChecksum}{\vars{fileSize}}
    \State MessageDigest\vars{~md}
	\State $\vars{totalBytes} = 0$
	\While {\vars{totalBytes} $<$ \vars{fileSize}}
	    \State \vars{buffer} $=$ \vars{queue}.\func{remove}()   \Comment Synchronized operation
	    \State \vars{md}.\func{update}(\vars{buffer})
	    \State \vars{totalBytes} += \vars{buffer.length}
	 \EndWhile
	 \State \vars{checksum} $=$ \vars{md}.\func{digest()}
	 \State \vars{socket}.\func{write}(\vars{checksum}) \label{line:sendChecksum}
	 \Return 
\EndFunction
\end{algorithmic}
\label{alg:receiver}
\end{algorithm}

Algorithm~\ref{alg:sender} and~\ref{alg:receiver} illustrate how \algoName~operates. Two threads are running simultaneously; one reads the file and sends it over the network and the other one computes the checksum. Since we observed that files that are smaller than free memory space are causing checksum I/O to be served from cache memory, \algoName~reads files once and shares data between threads via a synchronized $queue$ as shown in line~\ref{line:add} in Algorithm~\ref{alg:sender} and Algorithm~\ref{alg:receiver} to eliminate unnecessary system calls. Once file transfer is completed and both sender and receiver calculate the checksum of the file, receiver sends its own copy to sender to compare against source's (line~\ref{line:sendChecksum} in Algorithm~\ref{alg:receiver}). If checksum values do not match, then destination copy of the file is assumed to be corrupted and file is being transferred again. The execution time of this process equals to the slowest of transfer and checksum computation operations as it executes them simultaneously and waits for both to finish. $queue$ has fixed limited size, so even if transfer speed is much faster than checksum speed, it will not cause the queue to grow too large. If transfer operation is faster and $queue$ is filled, then transfer operations will need back-off run at at same speed as checksum computation. Although multiple CPU cores can be used to speed up checksum computation and multiple network parallel streams can be opened enhance transfer speed, the focus of this paper is not optimization of network transfer or checksum computation. Rather, it is optimization of their execution orders and I/O accesses to achieve an execution time that is closer to the slowest of these operations when they were run in isolation.

\section{Evaluations} \label{sec:results}
We tested \algoName~in ESNet (Berkeley, CA) and HPCLab (Reno, Nevada) whose specifications are given in Table~\ref{tab:esnet-lan} and~\ref{tab:system-spec}. ESNet servers are connected in two different paths, first one (ESNet-LAN) is through a top-of-rack switch and have 0.02 ms RTT. The second one (ESNet-WAN) is a loop from ESNet@Berkeley to Starlight@Chicago and back to ESNet@Berkeley whose RTT is 89 ms. For HPCLab experiments, we used two pairs of nodes; one pair involves two workstations (WS1 and WS2) in local area network with 1 Gbps bandwidth (HPCLab-1G) and the other one is comprised of two Data Transfer Nodes (DTN1 and DTN2) with 40 Gbps connectivity in local area network (HPCLab-40G). To enrich test scenarios, we added 30 ms emulated delays between these DTNs. We report the results in terms of overhead which is calculated as percentage of increase in time induced algorithms compared to time of the slower of transfer and checksum processes as shown in Equation~\ref{eq:overhead}. $t_{chksum}, t_{transfer}, t_{algorithm}$ refer to time it takes to run checksum computation, file transfer, and an integrity verification-enabled file transfer algorithm, respectively. For example, if file transfer without integrity verification takes 90 seconds, checksum computation takes 120 seconds, and \algoName~runs 130 seconds, then the overhead becomes $\frac{130-max(120,90)}{max(120,90)} = 8.3\%$. We also collected hit ratios to compare disk access behavior of the algorithms. Hit ratio is calculated as proportion of page accesses that are found in page cache (stored on main memory) to all page accesses. For instance, if 100 pages of a file are requested and 80 of them are found in page cache and remaining 20 are fetched from disk, then hit ratio becomes 80\%.

\begin{equation}
\frac{t_{algorithm} - max(t_{chksum}, t_{transfer})}{max(t_{chksum}, t_{transfer)}}
\label{eq:overhead}
\end{equation}

\begin{table*}[ht]
\begin{centering}
\begin{tabularx}{\textwidth}{ @{}|Y|Y|Y|@{}}
\hline
 {\bf Specs} & {\bf WS1 - WS2 (HPCLab-1G)} & {\bf DTN1 - DTN2 (HPCLab-40G)}\\
\hline
{\bf File System}  & Direct-Attached HDD & Direct Attached NVMe SSD\\
\hline
{\bf CPU} & 8 x Intel Core i5-7600 @3.50GHz & 16 x Intel Xeon E5-2623 @2.60GHz\\
\hline
{\bf Memory Size (GB)}  & 16 \& 24& 64  \\
\hline
{\bf Bandwidth (Gbps)} &  1 & 40 \\
\hline
{\bf RTT (ms)} & 0.2 & 30 (Emulated)  \\
\hline
\end{tabularx}
\caption{System specifications of HPCLab network. WS1-WS2 pair is referred as HPCLab-1G and DTN1-DTN2 pair is referred as HPCLab-40G in the the rest of the paper.}
\label{tab:system-spec}
\end{centering}
\end{table*}

\begin{figure*}
\begin{center}
\includegraphics[keepaspectratio=true,angle=0,width=\textwidth] {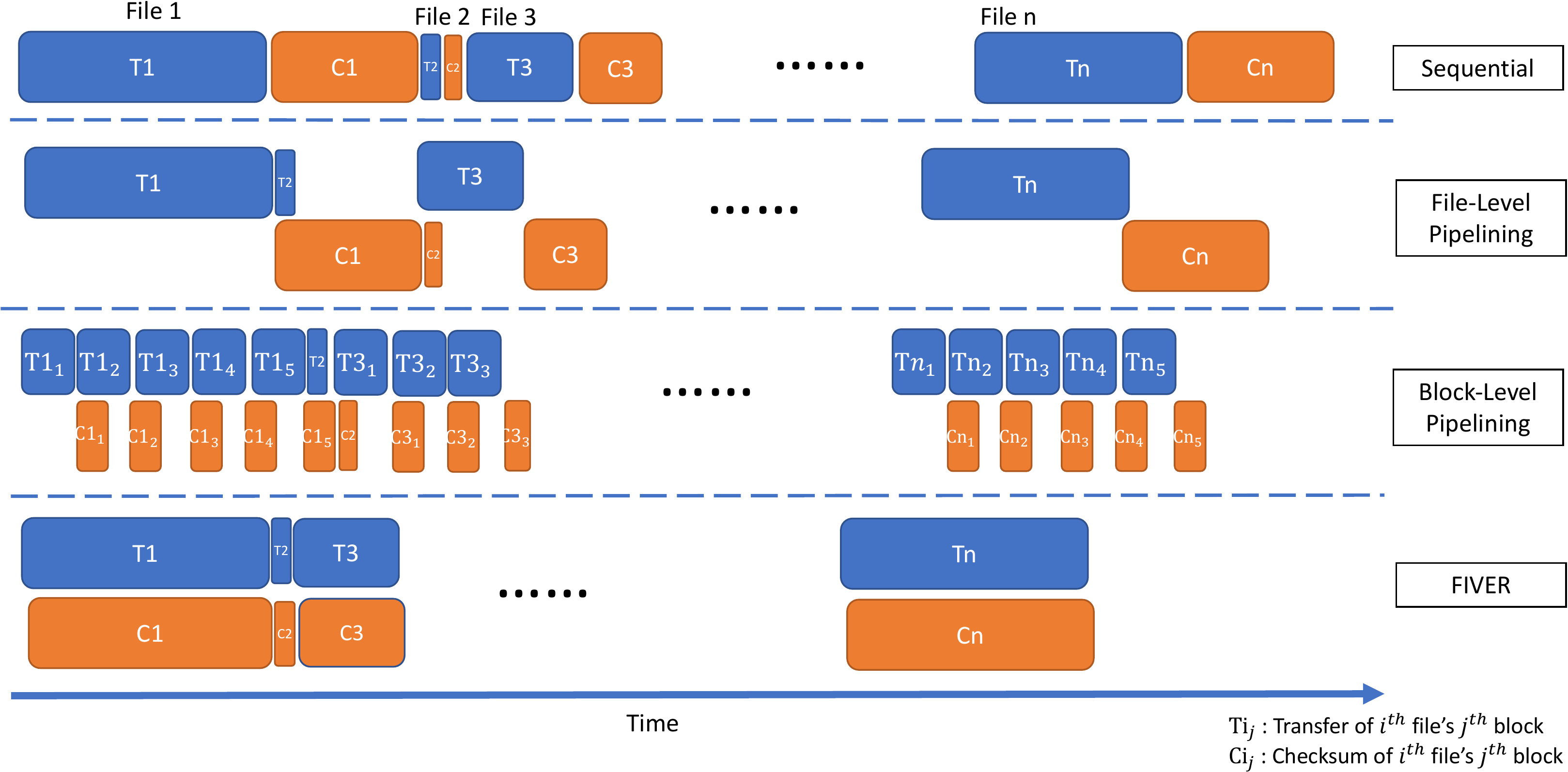}
\caption{Illustration of the order of transfer and checksum operations in different integrity verification implementations.}
\label{fig:illustration}
\vspace{-4mm}
\end{center}
\end{figure*}

We run the experiments using two types of dataset; uniform and mixed datasets. Uniform datasets consist of one or more files in same size. We created six uniform datasets where sizes of files are chosen to represent small and large files in each network. We created two types of mixed datasets: \emph{Shuffled} mixed dataset contains both small and large files and files are shuffled before the transfer to guarantee randomness in the order of transferred files. Example of a mixed dataset used in ESNet experiments is as follows: 100x10MB, 100x50MB, 50x250MB, 10x2GB, 4x8GB, 4x10GB, 1x15GB, and 2x20GB; in total of 271 files with total size 165.5GB. \emph{Sorted-5M250M} consists of equal number of 5M and 250M files that are arranged in a way that each 5M file is followed by a 250M file. The results are average of five runs.

We compare \algoName~against following algorithms:
\begin{itemize}
\item \textbf{Sequential}: Transfer and checksum operations are executed sequentially for every file. At any time, either file transfer or checksum operation runs but not both.
\item \textbf{File-Level Pipelining}: Transfer of a file is overlapped with checksum calculation of another file.
\item \textbf{Block-Level Pipelining}: Large files are split into small blocks of size 256 MB and checksum calculation of a block is overlapped with transfer of another block.
\end{itemize}

Figure~\ref{fig:illustration} depicts the order of operations for different integrity-verification algorithms. The first approach is sequential which runs file transfer and checksum computation operations in order. As a result, the total execution time equals to the sum of transfer time and checksum computation time of all files. Since there is a delay between consecutive file transfers, the transfer performance degrades as a result of TCP resetting its windows size which affects the performance considerably when transferring too many small files in wide area networks. File-level pipelining overlaps the transfer of a file with a checksum computation of another file. Block-level pipelining, on the other hand, tries to achieve better pipelining of transfer and checksum computation operations by dividing files into blocks. Finally, \algoName~runs transfer and checksum computation of each file simultaneously to benefit from I/O sharing and improve transfer performance.

\begin{figure*}
\begin{center}
\subfigure[Uniform Dataset]{
\includegraphics[keepaspectratio=true,angle=0,width=65mm] {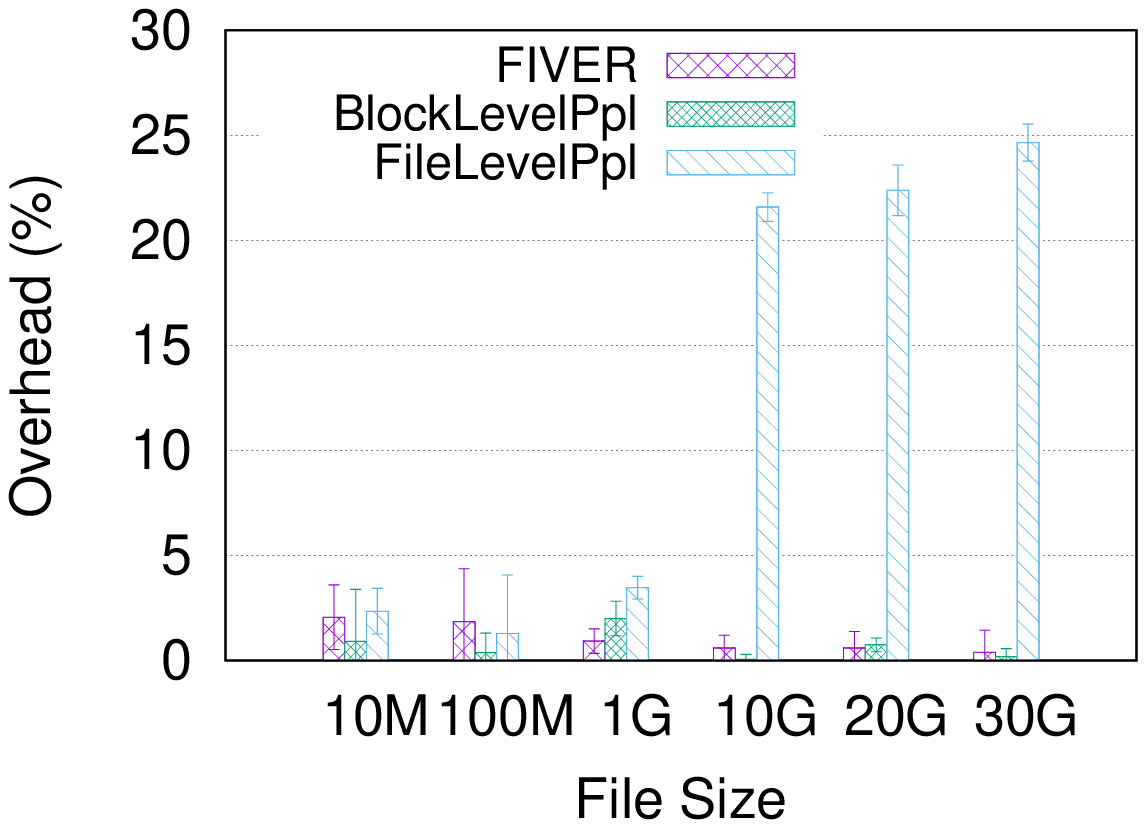}
\label{fig:local_single}}
\hspace{10mm}
\subfigure[Mixed Dataset]{
\includegraphics[keepaspectratio=true,angle=0,width=65mm] {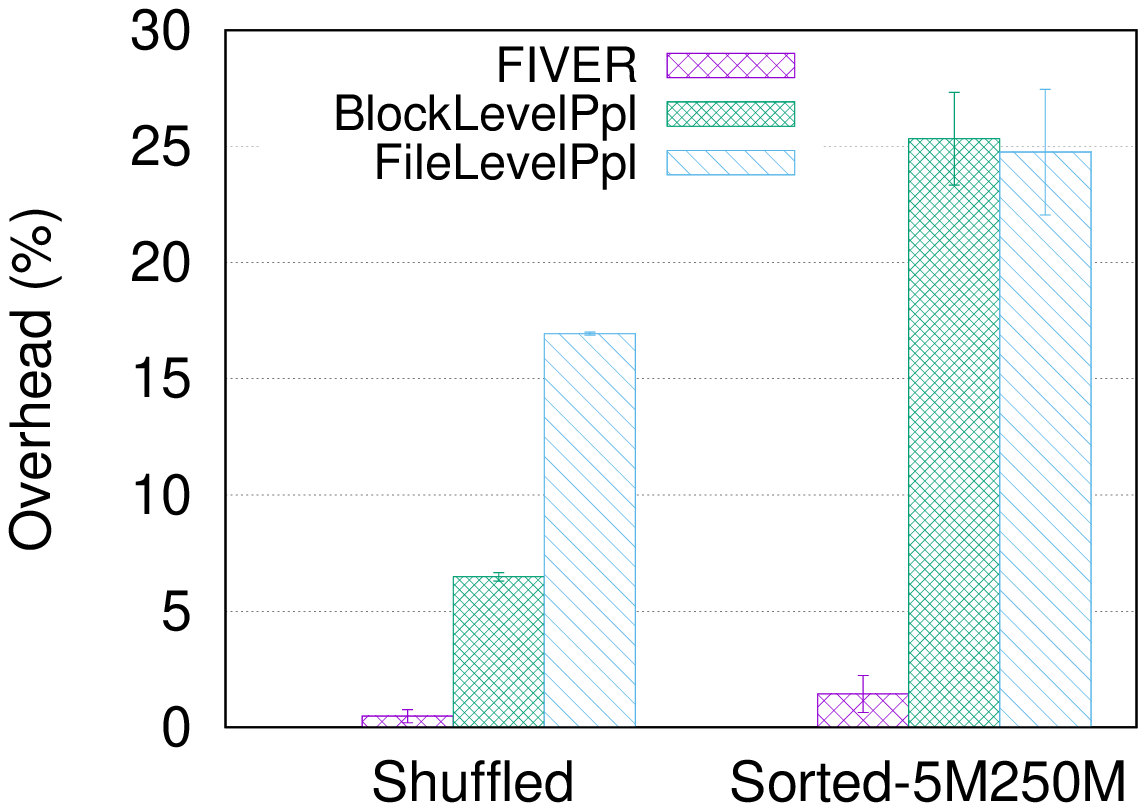}
\label{fig:local_mixed}}
\caption{Comparison of algorithms in HPCLab-1G network. The speed of checksum is faster than the speed of transfer.}
\label{fig:local}
\vspace{-3mm}
\end{center}
\end{figure*}

\begin{figure*}[ht]
 \centering
\includegraphics[keepaspectratio=true,angle=0,width=\textwidth]{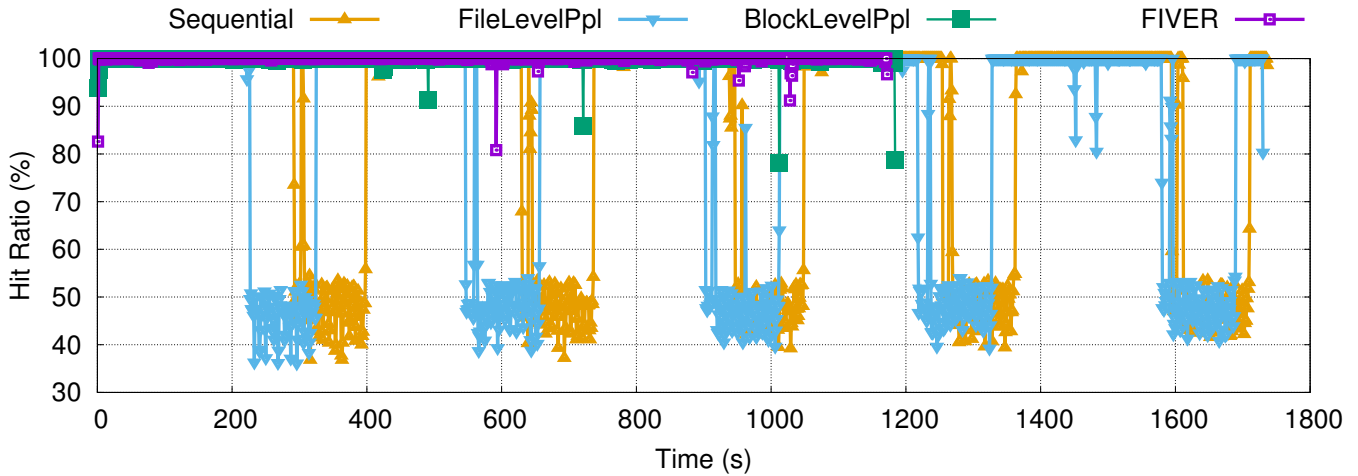}
\caption{Hit ratios of algorithms for mixed dataset transfer in HPCLab-1G.}
\label{fig:hpclab1G_hitratio}
\vspace{-4mm}
\end{figure*}

File-level pipelining generally achieves better performance than sequential approach by overlapping checksum and transfer computation of consecutive files. However, its performance is affected by the file size distribution of dataset as well as speed difference between transfer and checksum operations. For example, if files are ordered in a way that 10 MB file is followed by a 10 GB file, then it will overlap transfer of 10GB file with a checksum computation of 10 MB file which will decrease the benefit of pipelining since the transfer of 10 GB file will take much longer than checksum computation of 10 MB file. While block-level pipelining outperforms file-level pipelining by dividing large files into smaller blocks, it has two main disadvantages. First, while it improves the pipelining problem of transfer and checksum operations between consecutive files in large datasets, the issue may still appear for files that are smaller than block size. For example, if the block size is set to 500 MB, misalignment will still happen for a dataset in which 10 MB files are followed by 500 MB files similar to Sorted-5M250M dataset since it will create one block for every file in the dataset and pipeline blocks in different sizes.
Finding the optimal block size could be challenging since small blocks will suffer from poor transfer throughput and large blocks will cause suboptimal pipelining of transfer and checksum operations. Secondly, when files are divided into blocks, it may have negative impact on transfer throughput if network speed is faster than checksum speed since transfers will finish early and wait for checksum which may cause TCP to reset its window size due to idle time~\cite{RFC2581}. Similar behaviour will appear for \algoName~as well, however block-level pipelining will encounter more as a result of creating more transfer units with blocks. On the other hand, \algoName~is able to achieve near-optimal pipelining by initializing checksum computation as soon as transfer starts. On the other hand, if checksum computations is faster than transfer speed, it will just wait for data to be available, so its total CPU time will not change.

\begin{figure*} [!ht]
\begin{center}
\subfigure[Uniform Dataset]{
\includegraphics[keepaspectratio=true,angle=0,width=65mm] {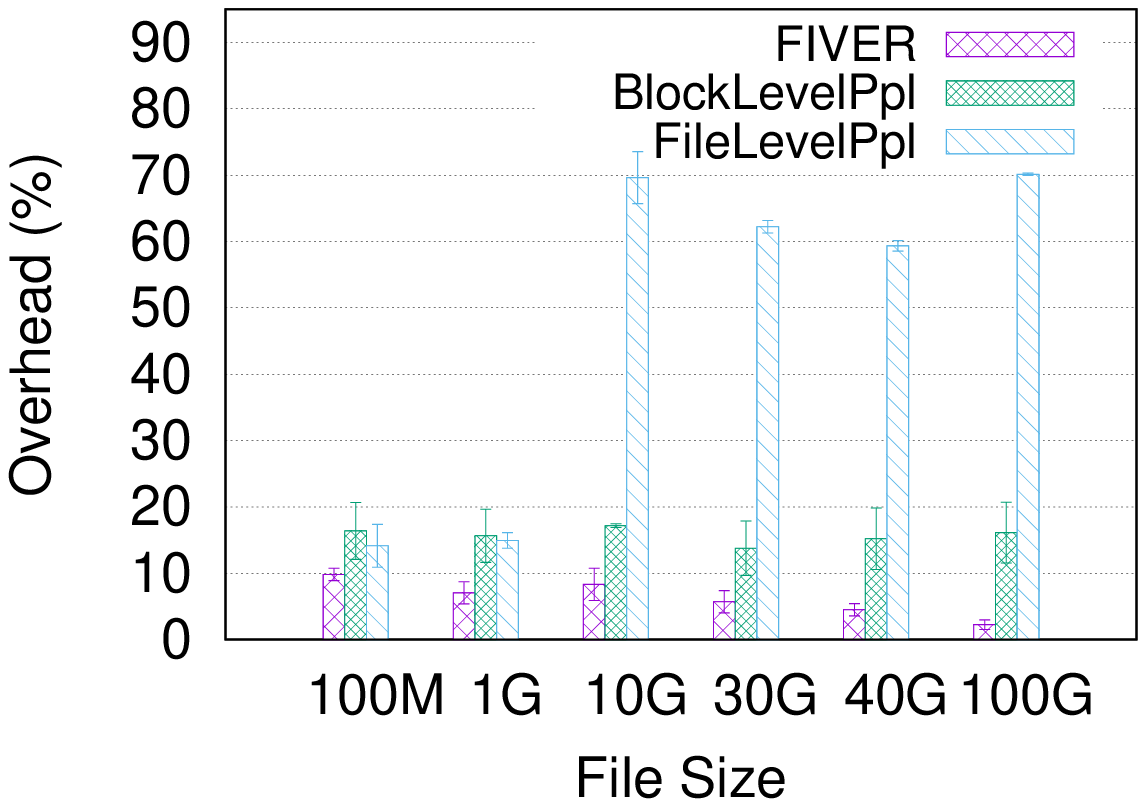}
\label{fig:dtn_uniform}}
\hspace{10mm}
\subfigure[Mixed Dataset]{
\includegraphics[keepaspectratio=true,angle=0,width=65mm] {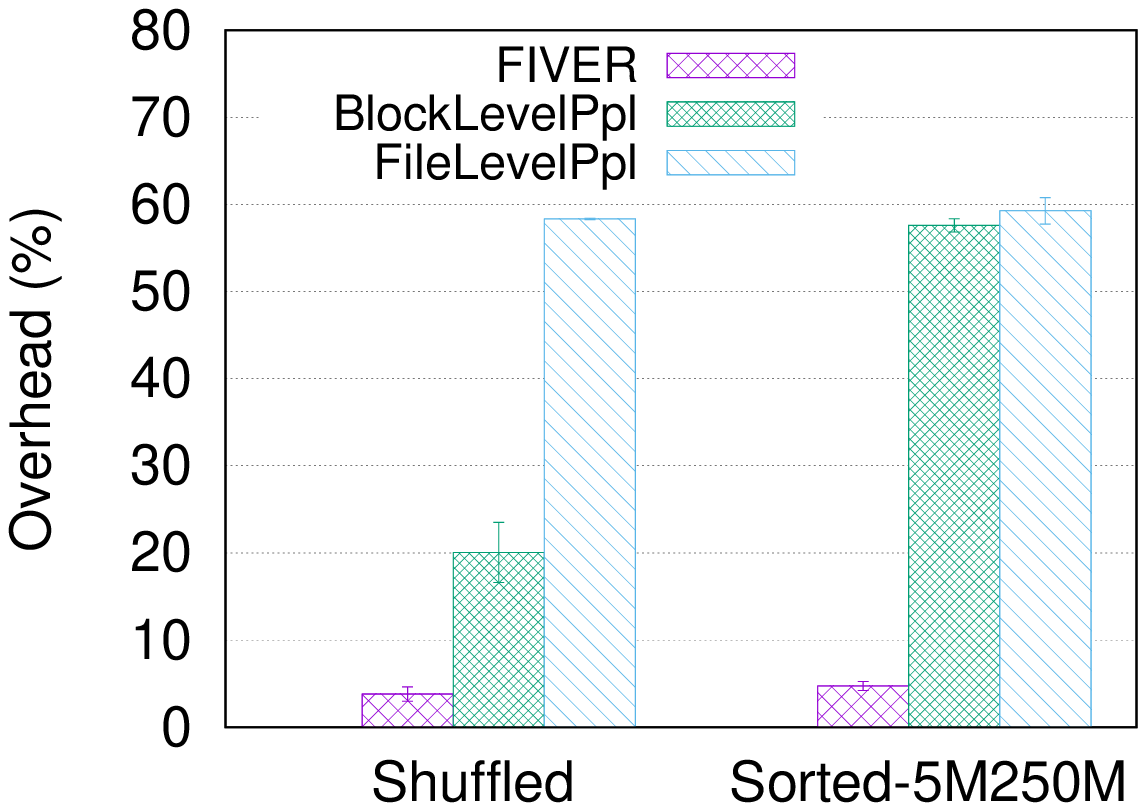}
\label{fig:dtn_mixed}}
\caption{Comparison of algorithms in HPCLab-40G network. The speed of transfer is faster than the speed of checksum.}
\label{fig:dtn}
\vspace{-4mm}
\end{center}
\end{figure*}

\begin{figure*}
\begin{center}
\subfigure[Uniform Dataset]{
\includegraphics[keepaspectratio=true,angle=0,width=65mm] {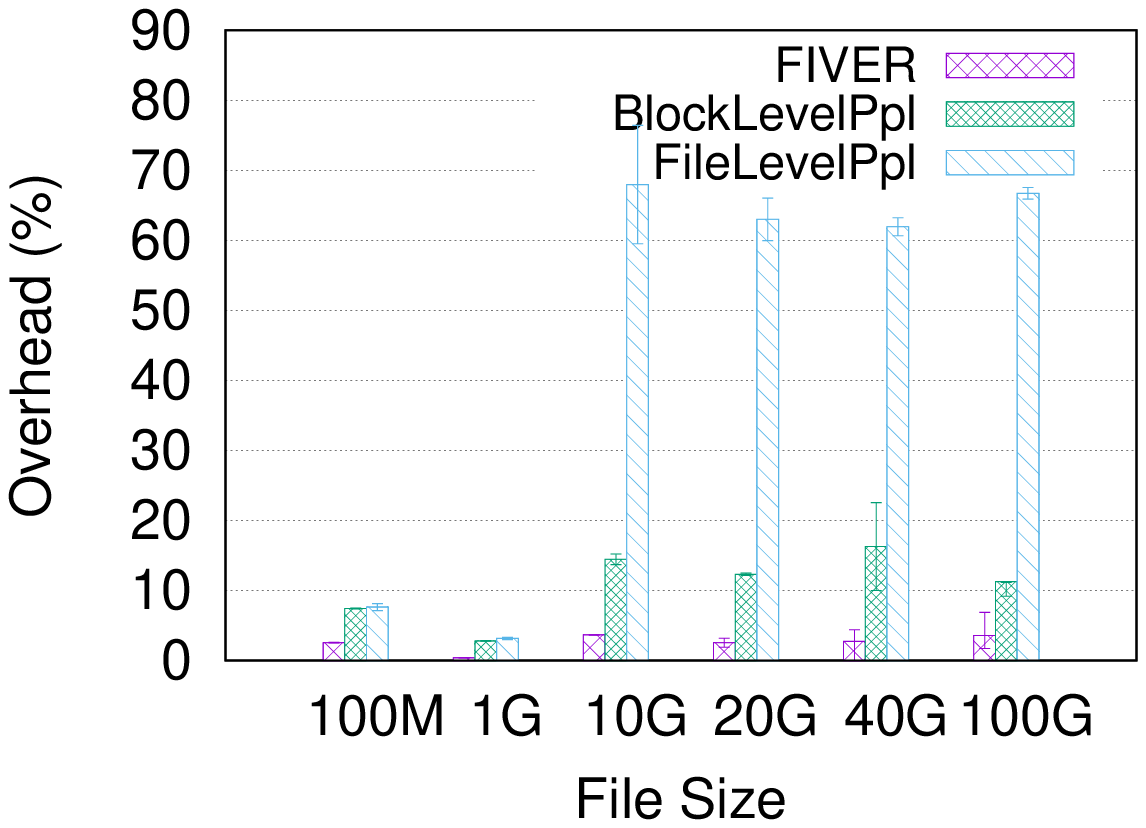}
\label{fig:esnet_small}}
\hspace{10mm}
\subfigure[Mixed Dataset]{
\includegraphics[keepaspectratio=true,angle=0,width=65mm] {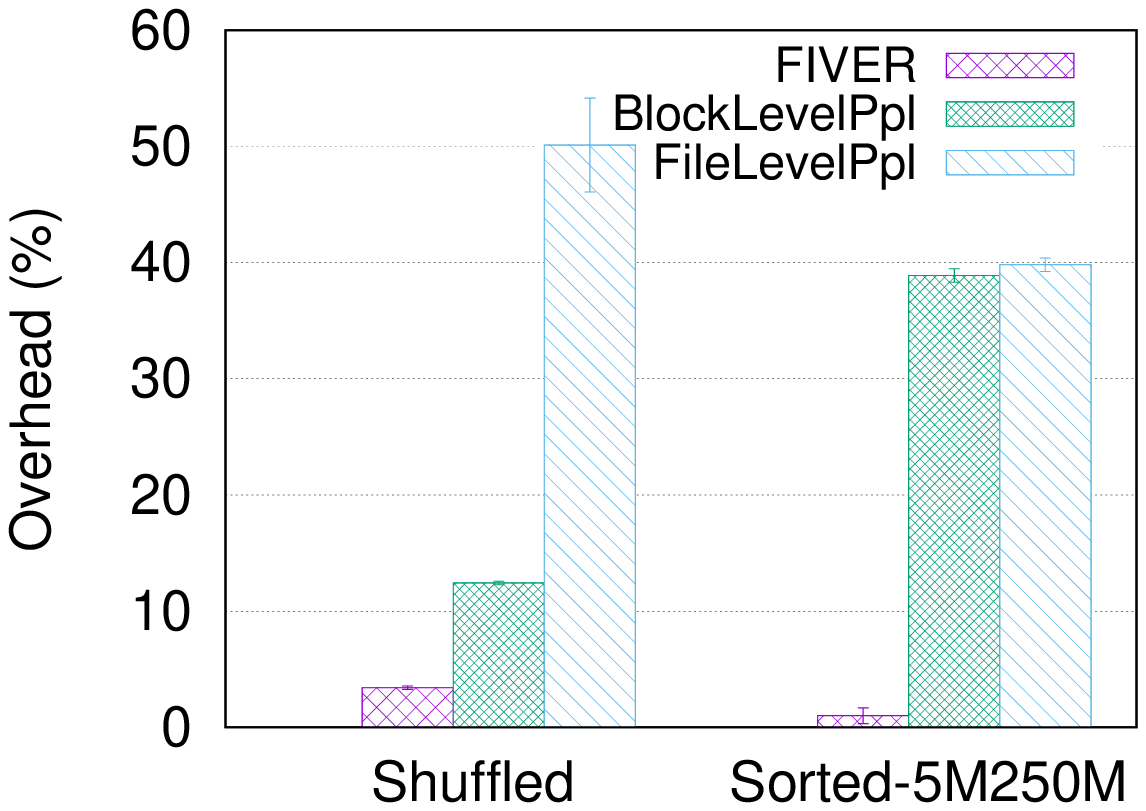}
\label{fig:esnet_mixed}}
\caption{Comparison of algorithms in ESNet-LAN network. The speed of transfer is faster than the speed of checksum.}
\label{fig:esnet}
\vspace{-5mm}
\end{center}
\end{figure*}

Figure~\ref{fig:local} shows the results of experiments in HPCLab-1G network in which network performance is slower than checksum calculation speed. Since \algoName~creates two threads, one for network transfer and one for checksum computation, the thread that handles file transfer determines the baseline for the overhead calculations as given in Equation~\ref{eq:overhead}. Figure~\ref{fig:local_single} shows the results for uniform datasets. While all algorithms perform similar for small files, the overhead of file-level pipelining (FileLevelPpl) increases up to 25\% for large files. This is merely because while there are multiple files in small datasets (e.g. 1000 files in 10M dataset), whereas there is only one file in large datasets, therefore the benefit of pipelining is not observed. Block-level pipelining (BlockLevelPpl) imposes similar overhead compared to \algoName~because checksum calculation is faster than transfer, letting network transfers run uninterrupted which prevents TCP window size resets that would negatively impact the performance of block-level pipelining. Moreover, both block-level and file-level pipelining algorithms are exposed to overhead increases for mixed dataset transfers (Figure~\ref{fig:local_mixed}) as a result of failure to sustain efficiency in pipelining. For example, Shuffled dataset contains 10 MB and 500 MB files and if two files from these groups are pipelined, then the benefit of pipelining would be marginal. Even though block-level pipelining splits up files into smaller blocks (256 MB in size), the misalignment will still appear in a smaller scale. While overhead ratio of block-level pipelining is 6\% for Shuffled dataset, it reaches to more-than 20\% for Sorted-5M250M dataset since it will not split 250M files into smaller blocks and perform similar to file-level pipelining which performs poorly due to pipelining 250M files with 5M files. On the other hand, \algoName~is able to keep the overhead less than 3\% for all uniform datasets and less than 1\% for both mixed datasets. 

Figure~\ref{fig:hpclab1G_hitratio} shows receiver-side hit ratio changes when Shuffled dataset is transferred using different approaches. It is clear that block-level pipelining leads to almost 100\% hit ratio throughout the transfer and integrity verification which corroborates the design principles of \algoName~that involves running network transfer and checksum computation of each file simultaneously and sharing of file I/O between them. Despite similarity in hit ratio behaviours, \algoName~is able to finish the integrity verification 25 seconds earlier than block-level pipelining. On the other hand, file-level pipelining and sequential approaches result in 84.1\% and 84.4\% average hit ratios since there are five 20GB files that are larger than free memory size (16 GB) which causes cache hit ratio to fall below 50\% during the checksum computation.

\begin{figure*}
\begin{center}
\subfigure[Uniform Dataset]{
\includegraphics[keepaspectratio=true,angle=0,width=65mm] {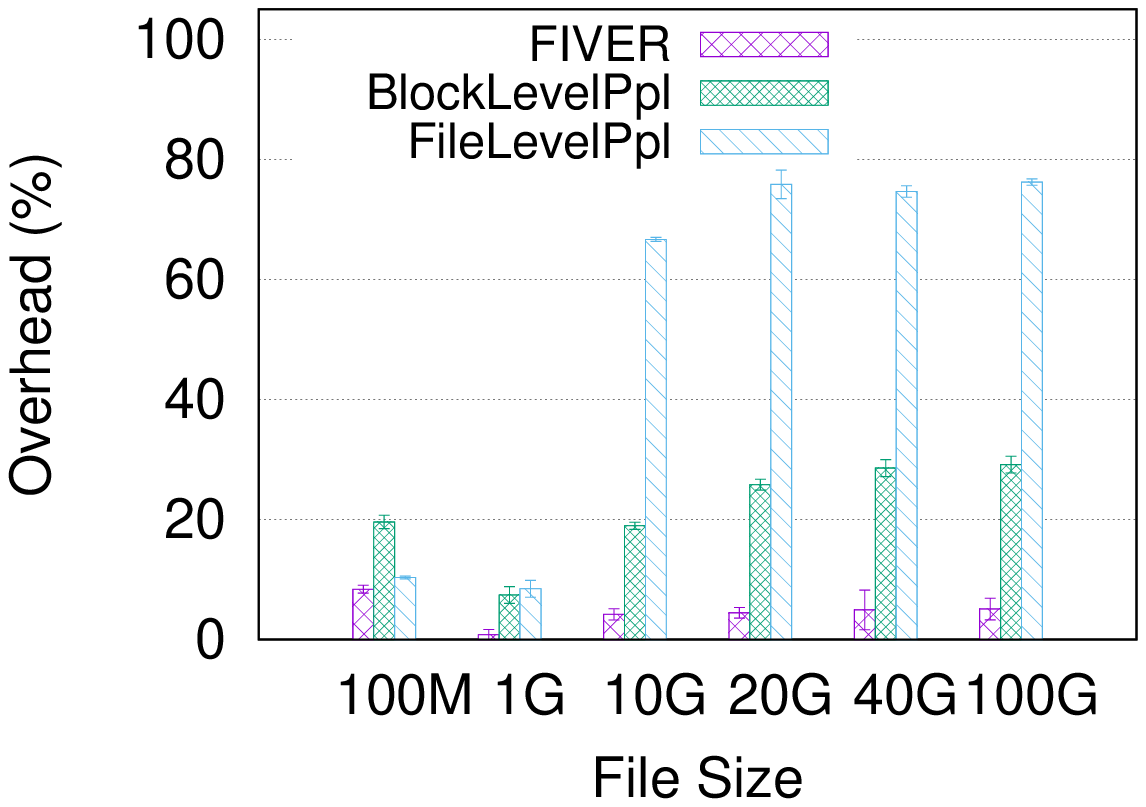}
\label{fig:esnet_wan_small}}
\hspace{10mm}
\subfigure[Mixed Dataset]{
\includegraphics[keepaspectratio=true,angle=0,width=65mm] {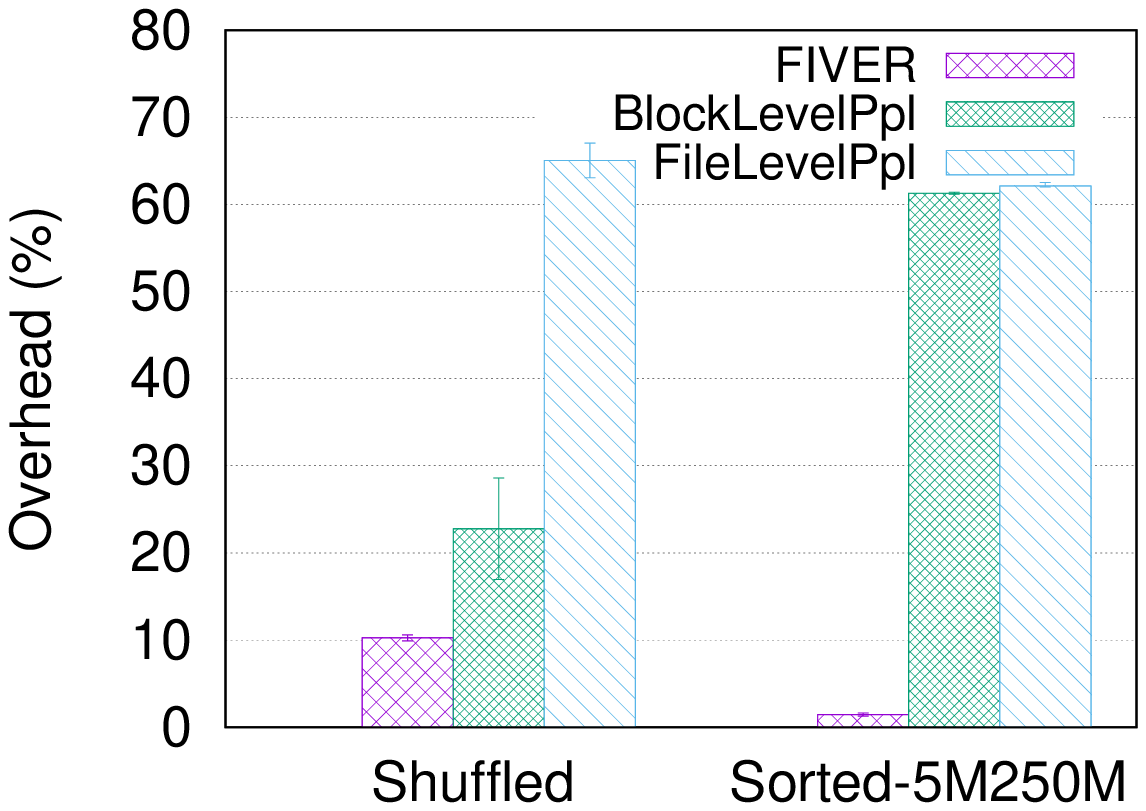}
\label{fig:esnet_wan_mixed}}
\caption{Comparison of algorithms in ESNet-WAN network. The speed of the network is faster than speed of checksum.}
\label{fig:esnet_wan}
\vspace{-4mm}
\end{center}
\end{figure*}

\begin{figure*}
\begin{center}
\includegraphics[keepaspectratio=true,angle=0,width=\textwidth] {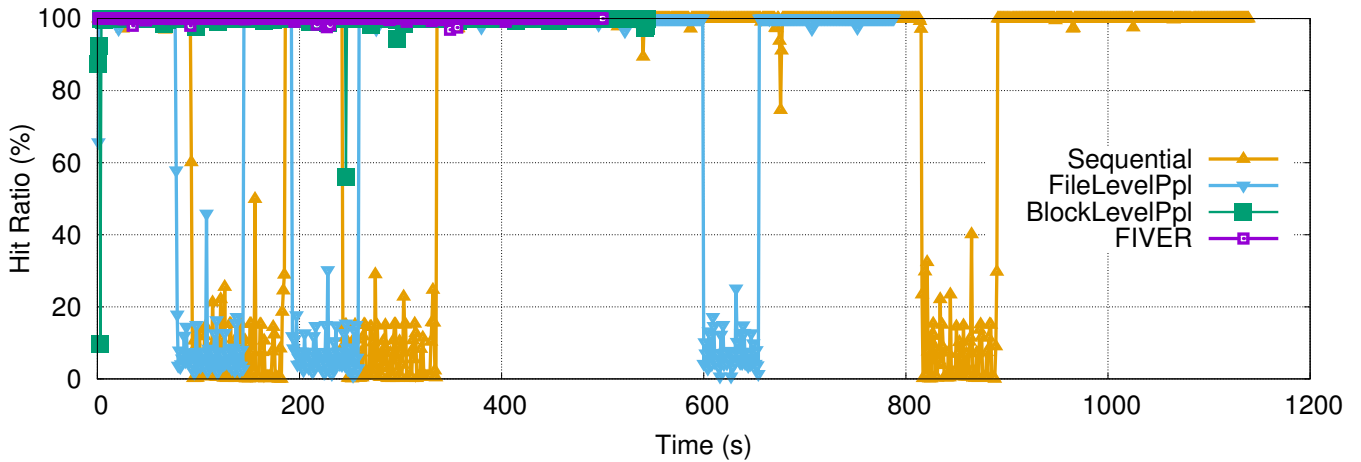}
\caption{Comparison of hit ratios for different algorithms for mixed dataset transfer in ESNet-WAN.}
\vspace{-4mm}
\label{fig:esnet_wan_hitratio}
\end{center}
\end{figure*}

Figure~\ref{fig:dtn} presents the overhead results in HPCLab-40G network where network speed is faster than checksum speed. As a result, \algoName~outperforms block and file-level pipelining algorithms by optimizing checksum calculation through file I/O sharing with network transfer. Block and file-level pipelining execute system calls to open and read files to calculate checksum which causes overhead because of context switching between user and kernel modes. Yet, as we explored in Figure~\ref{fig:hpclab1G_hitratio}, kernel fetches pages of files from the page cache for block-level pipelining. Thus, block-level pipelining observes up to 13-16\% overhead while~\algoName~can sustain less than 10\% overhead for all file types in uniform dataset. Overhead of file-level pipelining can go up to 70\% because it fails to benefit from pipelining when there is only one file in the dataset. Its overhead is less than 20\% for uniform datasets with multiple files (100M and 1G file types). Figure~\ref{fig:dtn_mixed} shows the overhead ratios when mixed datasets are transferred with different algorithms. Unlike uniform dataset, block-level pipelining observes higher overhead ratios due to presence of files that are smaller than block size which causes suboptimal transfer and checksum pipelining. Its overhead is 20\% and around 60\% for Shuffled and Sorted-5M250M datasets. \algoName~is still able to keep the overhead less than 5\% for mixed datasets. File-level pipelining leads to 55-60\% overhead for both mixed datasets which can be attributed to its imperfect pipelining performance in mixed datasets.

Figure~\ref{fig:esnet} shows the results of experiments in ESNet-LAN network. Since the network bandwidth is 40 Gbps and the speed of checksum computation is around 3 Gbps, checksum computation becomes the bottleneck. For example, a 100G file is transferred in 140 seconds (disk I/O is limited to 5-6 Gbps) but it took 273 seconds to compute its checksum. \algoName~and block-level pipelining achieve less than 10\% overhead for small files, however overhead of block-level pipelining increases to around 15\% for large files. For mixed datasets, \algoName~is able to keep the overhead less than 5\% similar to HPCLab results. The overhead of block-level pipelining is 12\% and 38\% for Shuffled and Sorted-5M250M datasets, respectively. File-level pipelining incurs 52\% and 39\% overhead to Shuffled and Sorted-5M250M dataset transfers.

\begin{figure*} [ht]
\begin{center}
\includegraphics[keepaspectratio=true,angle=0,width=\textwidth] {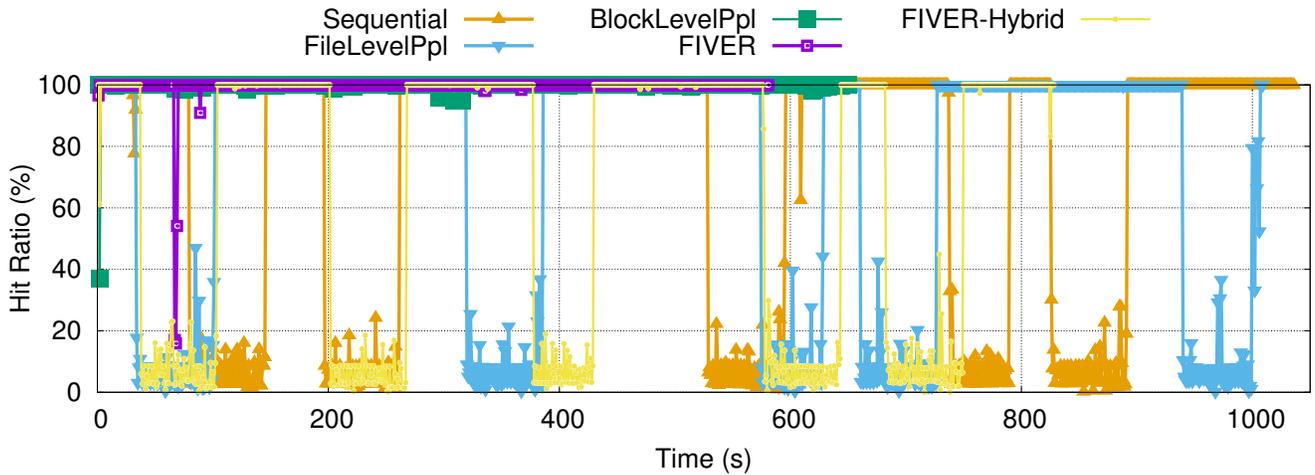}
\caption{Hit ratio analysis of \algoName-Hybrid for mixed dataset transfer in ESNet-WAN. It reduces execution time by 20\% compared to file-level and sequential algorithms while achieving similar cache access pattern.}
\vspace{-4mm}
\label{fig:esnet_wan_hitratio_hybrid}
\end{center}
\end{figure*}

We also tested the algorithms in the long path of ESNet network (ESNet-WAN) with 89 ms RTT. Similar to ESNet-LAN experiments, \algoName~achieves less than 10\% overhead for all file types whereas the overhead of block-level pipelining is around 15\% for uniform datasets and 20\% and 61\% for mixed datasets as shown in Figure~\ref{fig:esnet_wan}. As transfers last longer in wide area networks, overhead ratios increased a bit. For example, block-level pipelining is led to 20\% overhead for Shuffled dataset whereas its overhead was 12\% in ESNet-LAN. Similarly, file-level pipelining causes over 60\% overhead in ESNet-WAN network unlike 50-55\% overhead in ESNet-LAN. Moreover, block-level pipelining causes up to 60\% overhead for Sorted-5M250M dataset as shown in Figure~\ref{fig:esnet_wan_mixed}, again due to pipelining 5M blocks with 250M blocks that results in poor pipelining outcome. Figure~\ref{fig:esnet_wan_hitratio} depicts the hit ratios for mixed (Shuffled) dataset transfer. Again, cache hit ratio of \algoName~and block-level pipelining are almost always 100\% with 99.96\% and 99.69\% average values, respectively. On the other hand, \algoName~finishes 50 seconds earlier than block-level pipelining with the help of better pipelining of checksum and transfer operations as well as eliminating system calls in checksum computation. Moreover, file-level pipelining and sequential approaches experience time periods during which hit ratio falls below 10\% with average of 78.5\% and 77.8\%, respectively. Otherwise, their hit ratio is also almost always 100\% which can be attributed to the fact that kernel serves checksum I/O requests from cache memory for all algorithms if file size is smaller than free memory space.

\iffalse
\begin{figure*}[t]
\begin{center}
\subfigure[Small Files]{
\includegraphics[keepaspectratio=true,angle=0,width=58mm] {figures/xsede/small}
\label{fig:xsede_small}}
\hspace{-4mm}
\subfigure[Large Files]{
\includegraphics[keepaspectratio=true,angle=0,width=58mm] {figures/xsede/large}
\label{fig:xsede_large}}
\hspace{-4mm}
\subfigure[Mixed Files]{
\includegraphics[keepaspectratio=true,angle=0,width=58mm] {figures/xsede/mixed-crop.pdf}
\label{fig:xsede_mixed}}
\caption{Comparison of algorithms in XSEDE network. The speed of network transfer is slower than the speed of checksum computation.}
\label{fig:xsede}
\end{center}
\end{figure*}
Finally, we ran tests in XSEDE using XStream and Stampede2 sites. The speed of the network is limited to 1 Gbps due to which transfer operations become the bottleneck as can be seen in Figure~\ref{fig:xsede}. Similar to ESNET (WAN), splitting large files into many small blocks negatively affects the performance of Block-level pipelining. Taking {\it Transfer Only} as a baseline, \algoName~cuts the integrity verification overhead by 77\%, 60\%, and 78\% compared to Block-level pipelining for 100x100M, 10X1G and 1x10G files as shown in Figure~\ref{fig:xsede_small}. It also shortens the checksum overhead by 89-93\% for large files compared to Block-level pipelining. Furthermore, it reduces total execution time by 39\%, 53\% and 51\% for large files compared to File-level pipelining. \algoName~also shortens execution time nearly 10\% compared to pipelining based solutions for mixed dataset as shown in Figure~\ref{fig:xsede_mixed}.
\fi

\subsection{Efficient Error Recovery}
When integrity verification algorithms detect an unmatched checksum for a file/block, they need to transfer the file/block again to guarantee the integrity of the file/block at destination. For algorithms that run integrity verification at file-level, this means repeating transfer and checksum operations for whole file again even for a single bit failure which could increase execution time significantly especially when file size is too large. Block-level pipelining offers quick recovery under such circumstances by resending only small portion of the file since it operates on blocks where block size is chosen to be less than 1 GB. For example, if one bit of 100 GB file is corrupted during data transfer, then block-level pipelining will be able to recover the failure by resending only one block that contains the failed bit. On the other hand, the algorithms that checks the integrity at file level (file-level pipelining and \algoName) can also run integrity verification at sub-file resolution without degrading the performance. In fact, we implemented chunk-level integrity verification for \algoName~as follows: \algoName~keeps track of processed data size inside $computeChecksum()$ function in Algorithm~\ref{alg:sender} and \ref{alg:receiver} and when it  reaches to a certain value ($CHUNK\_SIZE$), receiver computes the checksum and sends it to sender to compare against corresponding block checksum. Since checksum computation is executed when $update()$ function is called, $digest()$ function call has negligible computational cost. Thus, frequent execution of $digest()$ function and checksum exchange operations does not affect the performance \algoName~too much unless $CHUNK\_SIZE$ is too small. If integrity verification fails for a chunk of a file, then the sender creates a new file with same metadata as the original file except offset and length and adds it to queue to be transferred again. Hence, only small portion of the file is resent to recover failures. Table~\ref{tab:chunk} shows execution times of \algoName~and block-level pipelining when a dataset with 15 large files (10 of 1GB files and 5 of 10GB files) is transferred in HPCLAB-40G network under different number of fault scenarios. We injected faults by flipping a random bit of randomly-chosen files during the transfer operation. $CHUNK\_SIZE$ is chosen to be same as size of a block in block-level pipelining (i.e. 256 MB). \algoName~with file-level integrity verification suffers significantly when faults happen since it has to transfer whole file again. Its execution time almost doubles in case of 24 integrity verification failures compared to no-failure case. On the other hand, chunk-level version of \algoName~is able to handle faults by only transferring a small portion of the file due to which its execution time and total transferred data size increase as much as block-level pipelining. Moreover, chunk-level integrity verification implementation of \algoName~performs very close to file-level implementation in no-failure case which shows its negligible impact on performance it the absence of failures.

\begin{table}
\begin{centering}
\resizebox{0.48\textwidth}{!}{
\begin{tabular}{ |c| c| c| c|}
\hline
{\bf Fault} & \multicolumn{2}  {c|}{\bf \algoName} & {\bf Block-level}\\
 {\bf Count} &  File Int. Ver.  & Chunk Int. Ver. & {\bf Pipelining}\\
\hline
{\bf 0} & 179.2s & 180.2s & 204.2s \\
\hline
{\bf 8} &  253.1s & 186.2s & 208.8s \\
\hline
{\bf 24} & 347.3s & 198.5s & 222.3s \\
\hline
\end{tabular}}
\caption{Execution times of algorithms when faults happen. \algoName's~chunk-level integrity verification offers efficient error recovery by resending only small portion of the file.}\label{tab:chunk}
\vspace{-3mm}
\end{centering}
\end{table}

\subsection{Hybrid Approach to Minimize Cache Hit Ratio for Checksum Calculation}
Since \algoName~overlaps checksum computation and data transfer operations, one may argue that it does not capture possible disk write errors due to always reading from memory. However, we showed that block-level pipelining also always reads from memory since it tends to select block size to be less than 1 GB which is much smaller than memory size of most production servers. Hence, the integrity verification of \algoName~is at least as reliable as block-level pipelining. 

On the other hand, while all four approaches in Figure~\ref{fig:illustration} read files from memory if the file size is smaller than available memory space (as can be also seen in Figure~\ref{fig:esnet_wan_hitratio}), sequential and file-level pipelining approaches read files from disk when file size is larger than free memory space since file pages will be evicted from cache as the transfer operation is going on. Thus, we propose \algoName-Hybrid which combines sequential algorithm and \algoName~in a way that it uses \algoName~for files that are smaller than memory size and sequential solution otherwise. Results show that \algoName-Hybrid can reduce execution time while achieving similar cache hit ratio compared to sequential checksum for datasets with mixed files sizes.

Figure~\ref{fig:esnet_wan_hitratio_hybrid} shows the destination-side hit ratio comparison when a mixed dataset is transferred in ESNet-WAN network. Hit ratio values for \algoName~and block-level pipelining are almost always 100\%, as a result they finish in 587 seconds and 658 seconds, respectively. On the other hand, \algoName-Hybrid, file-level pipelining, and sequential approaches experience low hit ratios for five times as there are five files in the dataset that are larger than memory size. Hence, they incur less than 5\% hit ratio during the checksum computation of these five files which roughly takes 65 seconds for each file. Moreover, they all lead to \~2.5M total cache misses which can be used to infer similarity in cache access behavior. Yet, while it takes 1021 and 1037 seconds for file-level pipelining and sequential approaches to finish the transfer, \algoName-Hybrid reduces the execution time by around 20\% and completes the transfer in 837 seconds. So, we can conclude that \algoName-Hybrid can be used as an alternative to file-level and sequential approaches to optimize transfer time while maintaining the same degree of reliability in integrity verification.

\subsection{Impact of Hash Algorithm on Execution Time of Integrity Verification}
\begin{figure}[t]
 \centering
\includegraphics[keepaspectratio=true,width=72mm]{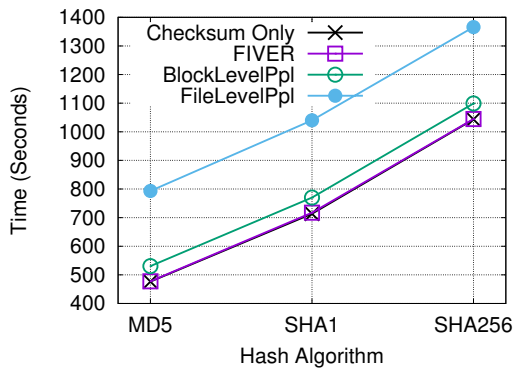}
\caption{Impact of hash algorithms on the execution time of different algorithms.} \label{fig:hash}
\vspace{-2mm}
\end{figure}

Although MD5 is still widely used, it has been found weak to collision attacks~\cite{wang2004collisions}. Hence, we compared execution times for MD5, SHA1 and SHA256 hash algorithms as shown in Figure~\ref{fig:hash}. We transferred the mixed dataset used in Figure~\ref{fig:esnet_mixed} in ESNet-LAN network. 

As expected, the time spent on checksum computation is proportional to the complexity of hash algorithm. For example, checksum computation without data transfer (Checksum Only) took 476, 713, and 1043 seconds for MD5, SHA1 and SHA256 algorithms, respectively. It took more than twice time to compute hash with SHA256 than it took using MD5. As a result, total execution time increases drastically for integrity verification algorithms, however \algoName~still imposes the lowest overhead compared to block-level and file-level pipelining. If we take Checksum Only as a baseline, block-level pipelining induced 50-60 seconds overhead whereas file-level pipelining imposed 300 seconds delay over baseline. It is also important to note that while total time increases, delay induced by different algorithms stays same as baseline values increase and transfer times do not change.

\section{Conclusion and Future Work} \label{sec:conclusion}
End-to-end integrity verification is vital for many applications which cannot tolerate silent data corruptions. However, it could degrade transfer performance significantly due to checksum computation which could take significant time. In this paper, we propose \algoName~and \algoName-Hybrid algorithms to minimize the cost of integrity verification. \algoName~overlaps the checksum computation and data transfer operations to reduce execution time and I/O overhead. The extensive results show that \algoName~ reduces the overhead of running integrity verification withing less than 10\% while existing approaches induces up to 60\% overhead. We further introduced \algoName-Hybrid that takes advantage of \algoName~for small files while following sequential approach for large files in an attempt to cut execution time without changing cache access behaviour of sequential approach. We showed that \algoName-Hybrid reduces total execution time by up to 20\% while having similar cache hit ratio. As a future work, we will examine ways to detect and solve possible data corruptions that may occur during memory-disk synchronization process such that, \algoName~can be used without deteriorating the reliability of integrity check.

\bibliographystyle{IEEEtran}
\footnotesize
\bibliography{references.bib}

\end{document}